\documentclass[USenglish,oneside,twocolumn]{article}

\usepackage[utf8]{inputenc}
\usepackage[big]{dgruyter_NEW}

\cclogo{\includegraphics{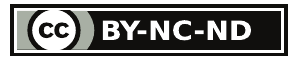}}

\PassOptionsToPackage{hyphens}{url}

\journalname{Proceedings on Privacy Enhancing Technologies}
\DOI{Editor to enter DOI}
\startpage{1}
\received{..}
\revised{..}
\accepted{..}

\journalyear{..}
\journalvolume{..}
\journalissue{..}

\usepackage{xspace}
\usepackage{graphics}
\usepackage{graphicx}
\usepackage{enumitem}
\usepackage{booktabs}
\usepackage{multirow, makecell}
\usepackage[binary-units=true]{siunitx}
\usepackage{amsfonts,amsthm,amsmath}
\usepackage{subcaption}
\usepackage{xcolor}
\usepackage{siunitx}
\usepackage{framed}
\usepackage{bm}
\usepackage{tabularx}
\usepackage{wasysym}
\usepackage{arydshln}
\usepackage{array}
\DeclareMathAlphabet{\mathsc}{OT1}{cmr}{m}{sc}
\makeatletter
\def\adl@drawiv#1#2#3{%
        \hskip.5\tabcolsep
        \xleaders#3{#2.5\@tempdimb #1{1}#2.5\@tempdimb}%
                #2\z@ plus1fil minus1fil\relax
        \hskip.5\tabcolsep}
\newcommand{\cdashlinelr}[1]{%
  \noalign{\vskip\aboverulesep
           \global\let\@dashdrawstore\adl@draw
           \global\let\adl@draw\adl@drawiv}
  \cdashline{#1}
  \noalign{\global\let\adl@draw\@dashdrawstore
           \vskip\belowrulesep}}
\makeatother

\usepackage{hyperref}

\newcommand*\dash{\ifvmode\quitvmode\else\unskip\kern.16667em\fi---%
\hskip.16667em\relax}

\newenvironment{sarahlist}{
\begin{description}[itemsep=2pt,leftmargin=0.4cm]
}{\end{description}}

\newcommand{\phead}[1]{\noindent{\bf\normalsize #1.}}

\newcommand{\ct}{Certificate Transparency\xspace}
\newcommand{\ctshort}{CT\xspace}
\newcommand{\caveat}{\dagger}

\begin{document}

\title{\huge SoK: SCT Auditing in \ct}

\author[1]{Sarah Meiklejohn}
\author[1]{Joe DeBlasio}
\author[1]{Devon O'Brien}
\author[1]{Chris Thompson}
\author[1]{Kevin Yeo}
\author[1]{Emily Stark}

\affil[1]{Google LLC, \{meiklejohn, jdeblasio, asymmetric, cthomp, kwlyeo, estark\}@google.com}

\begin{abstract}
{
The Web public key infrastructure is essential to providing secure
communication on the Internet today, and certificate authorities play a
crucial role in this ecosystem by issuing certificates.  These authorities may
misissue certificates or suffer misuse attacks, however, which has given rise
to the Certificate Transparency (CT) project.  The goal of CT is to store all
issued certificates in public logs, which can then be checked for
the presence of potentially misissued certificates.  Thus, the requirement
that a given certificate is indeed in one (or several) of these logs lies at
the core of CT.  In its current deployment, however, most individual
clients do not check that the certificates they see are in logs, as requesting
a proof of inclusion directly reveals the certificate and thus
creates the clear potential for a violation of that client's privacy.
In this paper, we explore the techniques that have been proposed for
privacy-preserving auditing of certificate inclusion, focusing on their
effectiveness, efficiency, and suitability in a near-term deployment.  In
doing so, we also explore the parallels with related problems involving browser
clients.
Guided by a set of constraints that we develop, we ultimately observe several key
limitations in many proposals,
ranging from their privacy provisions to the fact that they focus on the
interaction between a client and a log but leave open the question of how a
client could privately report any certificates that are missing.
}
\end{abstract}

\maketitle

\section{Introduction}

The basis for all secure communication on the Internet today is
the ability of a domain operator to
associate with their domain name a private key that they possess.  This is
achieved using digital certificates, which are issued by certificate
authorities (CAs) and which websites can present to clients, who can
check that they are valid by ensuring they are signed by, or have a signature
chain rooted in, a trusted CA.
If a CA is compromised, it can be used to issue false certificates
that in turn would allow an attacker to eavesdrop on the communication between
clients and a website.  Furthermore, CAs may simply fail to fully verify a
domain owner's identity and misissue a certificate.  Both of these scenarios
have happened~\cite{slate-article}, which illustrates the need for additional
mechanisms to avoid trusting CAs unconditionally.

It was out of this need that Certificate Transparency (CT) was created, with
the goal of increasing the visibility of certificates that
are issued and thus reducing the time required to detect misissued certificates
or other forms of CA misbehavior.  Briefly, this goal is achieved in CT by
requiring all issued certificates to be placed in one or more
public logs.
Upon receiving a certificate, a log operator checks its structural
validity and\dash assuming it is valid\dash responds with a \emph{signed certificate
timestamp} (SCT), which acts as a promise to include the certificate in the
log within some bounded delay.  These SCTs are then presented alongside the
certificate to the client, who checks that the SCTs are valid
and that there are enough SCTs according to the policy set by
their browser vendor.  Separately, designated
\emph{auditors} and \emph{monitors} are responsible for checking, respectively,
that logs are append-only and globally consistent
and that the actual content of the logs does not contain any inconsistencies or
misissued certificates.

In order for CT to achieve its goal, it is crucial that clients check not only
the validity of the SCTs they see, but also that the promises
implicit in these SCTs have been fulfilled; i.e., that these
certificates are in the log.  In this way, assuming auditors are doing
their job and everyone has access to the same log contents, clients can be
sure that the certificates they see will eventually be examined by a monitor.
If clients do not check that certificates are in the log, then even if they
were misissued they may never be detected by a monitor.  Having a client
check this directly with a log operator is a clear privacy violation, however,
as over time it amounts to having the client
present the log operator with their browsing history.  Thus, in the current
deployment of CT, most clients do not perform this crucial check.

\vspace{2mm}
\phead{Related work}
In 2013, Clark and van Oorschot systematically explored the space of known
security issues with HTTPS and evaluated proposals
for improving the certificate infrastructure in terms of their security and
privacy, deployability, and usability~\cite{clark13oakland}.  In 2020, Chuat et
al.\ explored the landscape of solutions to the problems of delegation and
revocation in the web public-key infrastructure~\cite{chuat20eurosp}.  Our paper
is complementary to these works: rather than perform a broad exploration
of secure communication on the Web and an evaluation of long-term
replacements or improvements, we seek to identify a pragmatic solution
to a specific problem, SCT auditing, that exists within the current HTTPS
ecosystem.  This could be used to further secure the ecosystem as it
exists today.

\vspace{2mm}
\phead{Our contributions}
Our contributions in this work are as follows:

\begin{itemize}[leftmargin=7pt]
  \item We survey the landscape of proposed solutions for the problem
    of privacy-preserving SCT auditing from the academic
    literature and the community of CT practitioners. We unify
    proposed solutions within a common framework of a
    \textit{querying} and \textit{reporting} phase.
  \item We compare SCT auditing to other privacy-sensitive
    problems in web browsers. We observe that
    while SCT auditing is a unique problem, some insights and
    approaches from other problems are applicable.
  \item We systematically evaluate proposed solutions on several
    dimensions: privacy, security, costs, and deployability.
  \item From our systematic evaluation, along with experimental data
    against real-world CT logs and insights from the deployment of CT
    and other related systems, we conclude that existing proposals
    suffer from crucial limitations. We extract a set of constraints
    that future SCT auditing solutions should adhere to, with the hope of
    guiding the community towards a complete solution.
\end{itemize}

\section{SCT Auditing}\label{sec:background}

Our description of \ct (\ctshort) and its components follows RFC~6962~\cite{6962}.
We refer at various points to specific aspects of another RFC,
6962-bis~\cite{6962bis}, which describes ``version 2.0 of the Certificate
Transparency protocol.''  To the best of our knowledge, RFC~6962 is an accurate
description of \ctshort as it exists today and there have been no announced
plans to migrate \ctshort to RFC~6962-bis.

\subsection{How \ct works}\label{sec:ct}

At a high level, \ctshort works as follows: a website operator requests a
\emph{certificate} from a CA, which should serve to bind together a domain
name and a public key.  \ctshort requires that every such certificate must
appear in at least one globally visible \emph{log}.  These logs are
append-only, which means that once a certificate has been added it cannot be
removed.  \ctshort thus exposes every certificate that has ever been issued.

The only restriction that logs place on the contents they store is that all
certificate chains must lead back to one of a set of trusted root CAs, published
by the log.  For almost all CAs, this means that the current advice is to
effectively broadcast a (pre)certificate to all log operators of whom they are
aware.  While certificates are typically submitted to logs by the CA at the time
they are issued, any other party can also do so at any time after the
certificate has been issued.

In order for a certificate to be valid, it should have at least one
\emph{signed certificate timestamp} (SCT) embedded into it, which proves that the
certificate has been submitted to a log and promises that it will be
included in the log within some \emph{maximum merge delay} (MMD).  Having the CA
wait only to receive this promise has several advantages over having them wait until
the certificate is actually included, such as allowing CAs to issue certificates
without relying on logs to perform a computationally intensive and potentially
time-consuming process; this freedom for logs to update less frequently
can also have a positive impact on privacy.  We discuss this in more
detail in Sections~\ref{sec:limit-latency} and~\ref{sec:limit-privacy}.

These SCTs can
be delivered to clients in three ways: (1) by embedding them in the certificate
directly using its extensions (which requires the CA to have submitted a
precertificate before issuance); (2) using a special TLS extension; and (3)
using OCSP (Online Certificate Status Protocol) stapling, which allows the CA to provide (and potentially obtain) the
SCT after the certificate has been issued.  Importantly, these three options and
the fact that certificates can be submitted to logs at any time means there is
no canonical set of SCTs associated with a given certificate.

Once a certificate has been included in the log, any party can request an
\emph{inclusion proof} from the log, which proves the certificate's inclusion
with respect to a timestamped commitment to the log's contents known as a
\emph{signed tree head} (STH).  Anyone can furthermore request a
\emph{consistency proof} between two STHs, which proves that the log operates in
an append-only manner.  A party in possession of some known STH may need to
obtain and verify both types of proofs in order to fully verify the inclusion of
an entry in the log: first an inclusion proof with respect to a given STH, and
then a consistency proof between this STH and their known one.  The only way
this second step would not be required is if the log operator happened to use the
party's known STH in their inclusion proof (e.g., it had not yet issued a new
STH for the log).

\subsection{The \ctshort ecosystem}

\begin{table}
\centering
\begin{tabular}{llS[table-format=4.1]S[table-format=4.1]S[table-format=3.1]S[table-format=2.2]}
Operator & Log name & {Log size} & \multicolumn{3}{c}{Log shard} \\
& & & {2021} & {2022} & {2023} \\
\midrule
Cloudflare & Nimbus & {--} & 408.9 & 382.2 & 6.7 \\
\multirow{2}{*}{DigiCert} & Nessie & {--} & 99.1 & 134.1 & 22.1 \\
    & Yeti & {--} & 305.3 & 66.5 & 22.2 \\
\multirow{6}{*}{Google} & Argon & {--} & 1356.2 & 671.4 & 25.5 \\
    & Xenon & {--} & 1464.4 & 833.8 & 26.3 \\
    & Icarus & 762.4 & {--} & {--} & {--} \\
    & Pilot & 1077.3 & {--} & {--} & {--} \\
    & Rocketeer & 1113.9 & {--} & {--} & {--} \\
    & Skydiver & 309.2 & {--} & {--} & {--} \\
Let's Encrypt & Oak & {--} & 365.3 & 242.9 & 20.6 \\
\multirow{2}{*}{Sectigo} & Sabre & 199.9 & {--} & {--} & {--} \\
    & Mammoth & 606.9 & {--} & {--} & {--} \\
TrustAsia & Log & {--} & 0.7 & 1.0 & 0.06 \\
\bottomrule
\end{tabular}
\caption{All usable \ctshort logs and their respective sizes, in terms of
millions of entries, as of February 23, 2022.  For the temporally sharded logs,
we present the sizes of the 2021, 2022, and 2023 shards.}
\label{tab:basics}
\end{table}

The issuance process described above almost always takes place between the CA
and the log, but there are other important participants in the \ctshort ecosystem.
All \ctshort-enforcing user agents (i.e., browser vendors) have their own policy in
terms of how many SCTs they require a certificate to have.
For example,
Apple (Safari) requires SCTs from at least two distinct log
operators and Google (Chrome) requires one SCT from a Google-operated log and one
from a non-Google log.

Table~\ref{tab:basics} contains a list of all currently usable \ctshort logs and
their
operators.\footnote{This list is taken from
\url{https://www.gstatic.com/ct/log_list/v3/log_list.json}.}
As we can see, there are currently billions of certificates stored in these
logs, and there are millions added every day~\cite{merkle-town}.  To control
the growth of any one log, new \ctshort logs are required to be \emph{temporally
sharded}~\cite{ct-log-policy},
with each log shard containing all certificates that expire within a given
time range (typically one calendar year).  Nevertheless, the sheer number of
certificates across all logs makes it difficult for an individual domain owner to
check for new (and possibly unauthorized) certificates being issued to their
domain.  In order to simplify this task, \emph{monitors} act as mirrors and
provide search and notification services~\cite{ct-monitors, li19ccs-ct}.
Some of the current prominent examples of monitors are crt.sh
(\url{https://crt.sh}) and Cert Spotter
(\url{https://sslmate.com/certspotter/}).
While monitors thus help detect misbehavior on the part of CAs,
\emph{auditors} are responsible for detecting misbehavior on the part of
logs.  This role can be played by any entity with the ability to report or act
on misbehavior; e.g., to initiate the process of removing a log from
the ecosystem~\cite{aviator-policy,aviator-policy2,startcom-policy,
startcom-policy2,wosign-policy,wosign-policy2}. Notably, individual browser
instances are not in a position to act as auditors, at least not without the
help of another entity such as the browser vendor.

In order to make sure that logs follow up on the promise implicit in an SCT, it
is necessary to check that the certificate represented by this SCT is in fact
included in the log, and act accordingly if not.  We refer to this process as
\emph{SCT auditing}.  Individual browser instances, however, are the
only participants who reliably see SCTs ``in the wild'', by
visiting websites and receiving certificates, and having them perform SCT
auditing is a clear privacy problem: an SCT uniquely identifies a certificate
and thus a domain (or set of domains if wildcards or alternative names are
used), so reveals the website visited by the browser.  Having
an individual browser instance reveal SCTs to another party, whether it is the
log or an auditor, thus reveals to that party some of the browsing history of
that user.  The problem of how to audit SCTs without compromising user privacy
has been open since the introduction of CT.

\subsection{Threat model}\label{sec:threat}

As described above, there are three core actors in an SCT auditing ecosystem:
(1) users, whose browsers see SCTs as they visit websites; (2) logs,
which store certificates and serve inclusion proofs for them; and (3)
auditors, who are responsible for acting on evidence of log misbehavior.
Ultimately, this means the goal of SCT auditing is for an honest auditor to learn
about any SCTs whose implicit promises have been violated.

We assume that users, auditors, and logs can all be malicious; i.e., can
attempt to deviate from the protocol in arbitrary ways.  Importantly though, we
assume that malicious parties cannot cause honest parties to deviate from the
protocol; e.g., if the auditor is a browser vendor they cannot inject
malicious code into the browser to affect the behavior of an honest user.
We also assume that all log operators use secure cryptographic standards, 
means they cannot form valid inclusion or consistency proofs unless,
respectively, an entry really is in the log or the log really is append-only.

We break the problem of SCT auditing into two phases: \emph{querying} and
\emph{reporting}.  We model each phase as an interaction
between a client and a server.  In the querying phase, the server is the
log operator, and the client's goal is to learn whether or not a specific entry
is included in the log.  In the reporting phase, the server is an
auditor, and the goal is for them to learn about any
entries that were not included in the log.

In both phases, the client is intentionally left generic, allowing it to represent
individual browser instances but potentially other participants (e.g., auditors or
web servers) as well.  In terms of privacy, the information in a query or
report\dash meaning the certificate or SCT it contains\dash is not
sensitive; i.e., it does not inherently leak anything about individual
clients.  In either phase, the goal of an adversary is thus to either link
together two queries (i.e., to learn that they came from the same user) or to
link an individual user to a specific certificate; i.e., to learn that they
queried on or reported a specific certificate.

\section{Related Problems}

We consider three problems that are related to the problem of SCT
auditing, in terms of providing protection to users as they browse the Internet.

\subsection{Safe Browsing}\label{sec:safe-browsing}

In order to protect users from phishing, their browsers
can periodically check whether or not the sites they visit are on a blocklist
maintained by Google.  The Safe Browsing API~\cite{sb-apis} supports two types
of interactions: basic lookups and updates.  A basic lookup reveals
the queried URL in the clear, and is analogous to a CT log's API for fetching
inclusion proofs.  The \texttt{Update} call allows clients to perform a local
lookup before deciding whether or not to interact with a Safe Browsing endpoint
directly.  Briefly, clients can store
the hash prefixes of URLs on the Safe Browsing list in a
compressed data structure similar to a Bloom filter.  When a client visits a
URL, they hash it and check if its hash prefix is in the filter or not.  If
not, then the URL is definitely not on the Safe Browsing blocklist (or at least
the version of it reflected by the client's filter).  If it is,
then they call the API on the hash prefix to get back a list of (full) hashes
of all URLs on the blocklist that have that prefix.  If the hash of the URL is
on that list then it is on the blocklist (and thus considered unsafe), and
otherwise it is not.

Currently, many browser vendors integrate with Safe Browsing, meaning their
users are by default opted in to use the \texttt{Update} API and update their
local filter every 30 minutes (they also have the ability to opt out).  Users of
Chrome may additionally opt in to ``enhanced'' Safe Browsing~\cite{esb-blog},
which means their browsers query the lookup API in real time.

The problem that Safe Browsing is solving is in some sense the inverse of the
problem in SCT auditing: in Safe Browsing users are looking for a match in a
relatively small list of URLs (the blocklist), whereas in \ctshort users are
looking for a missing entry in a large list (the contents of a CT log).  As we
explore further in Section~\ref{sec:psm-report}, this means the approach
used in Safe Browsing cannot be used directly in \ctshort.

\subsection{Checking for certification revocation}\label{sec:revocation}

When verifying a certificate, it is important to
ensure that it has not been
\emph{revoked}.  One way to check for revocation is using the Online Certificate
Status Protocol (OCSP).  This allows CAs to tell clients the status of a
certificate
via an endpoint that clients query
directly.  Using OCSP directly thus presents a privacy issue, as clients reveal
the certificates they see to the CA.

There are two basic alternatives to OCSP that exist
today.  Using OCSP Stapling, the certificate holder
(i.e., the web server) queries the CA rather than the client, and then
``staples'' the signed response from the CA to the certificate when it serves
it.
Since the web server can perform these queries periodically and
cache the response until some expiration date, this improves not only
privacy but also performance.  In the other alternative, clients can query
endpoints maintained by CAs to download certificate revocation lists (CRLs).
Clients can then check certificates against locally cached versions of these
CRLs, but this imposes a high storage overhead and
requires clients to keep their lists up-to-date.  To address these limitations,
CRLite~\cite{larisch17oakland,firefox-crlite} stores revocation data in a
compressed data structure similar to a Bloom filter, while CRLSets~\cite{crlsets}
are revocation lists of size at most 250\si{\kilo\byte} that are curated by Google
and pushed regularly to Chrome browsers.

As compared with SCT auditing, certificate revocation checks are designed to be
performed in-band during the setup of a connection, whereas \ctshort is designed
to have asynchronous auditing.  Furthermore, it is possible to have a
canonical list of all revoked certificates but not possible to enumerate all
SCTs/certificates that are not included in a \ctshort log.

\subsection{Checking for compromised credentials}

Users who are concerned that their
credentials may have been compromised can query and check whether or not these
credentials are on a list of breached credentials, as provided by a service such
as Google's Password Checkup~\cite{thomas19usenix} or Have I Been
Pwned? (HIBP)~\cite{hibp}.  While this is not directly related to browsing,
many browsers have integrated some form of checking; e.g., the Firefox
Monitor~\cite{firefox-monitor} uses
HIBP, while Safari and Chrome use their own custom lists.

Perhaps surprisingly, this problem is the one that most closely resembles SCT
auditing, as it also involves asynchronous querying in a large database.
Nevertheless, there are also important
differences, such as users needing to act in the case of matching rather than
missing data (as with Safe Browsing).  We discuss existing protocols for
checking for compromised credentials (C3) and how
they can be adapted for use in \ctshort in more
detail in Section~\ref{sec:psm-query}.

\section{Components of SCT Auditing}\label{sec:components}

\subsection{Literature review}

Our goal is to identify proposed solutions for either phase of SCT
auditing.  Given that \ctshort is a deployed project as well as an area of
academic research, we identified solutions based on a manual review of three
different sources: (1) academic literature
published in computer security and networking conferences (ACM CCS, USENIX
Security, IEEE S\&P, NDSS, NSDI, and CNS), (2) experimental deployments in
industry, and (3) posts on Certificate Transparency mailing
lists~\cite{ct-mailinglists} and
standards documentation.  The final list of proposals was compiled and
categorized jointly by a set of three researchers, and was validated by a
broader set of researchers and colleagues.
In addition to looking at proposals for \ctshort, we also looked at discussions
of and proposed solutions for the three related problems introduced in the
previous section.

\subsection{Evaluation criteria}

We consider seven main aspects that characterize a proposal for either phase
of SCT auditing.  A summary of all proposals against these evaluation criteria is
in Table~\ref{tab:query}.

\begin{sarahlist}

\item[Integrity:] We require that all proposals achieve integrity.  In the
querying phase, this means that the client accepts only entries that are
included in the log.  In the reporting phase, it means that the auditor
acts only on SCTs that have been violated.

\item[Privacy:] We consider whether or not a proposal preserves privacy.
Following Section~\ref{sec:threat}, this means that it is
difficult for the server to link a specific user (as
represented by their browser instance) to a specific entry.
We use $\Circle$ to indicate that no privacy is achieved;
$\LEFTcircle$ to indicate that $k$-anonymity is achieved, meaning the server
knows either that one of $k$ clients was interested in a specific entry or that
a specific client was interested in one of $k$ entries; and $\CIRCLE$
to indicate that unlinkability is provably achieved.

\item[Client costs:] We consider three costs that clients might incur:
bandwidth, storage, and computation.
We use $\Circle$ to indicate that there is no overhead; $\LEFTcircle$ to
indicate that there is some overhead but the client could still likely be run on
a modern mobile device; and $\CIRCLE$ to indicate that there is enough overhead
that it likely could not.

\item[Certificate issuance latency:] We consider any latency the protocol adds
to the process of certificate issuance.
We use `none' to
indicate that there is no added latency, and otherwise describe what is needed
before a certificate can be issued.

\item[Server costs:] We consider the costs for the server (i.e., the log or
auditor) to run the protocol.
We again use $\Circle$ to indicate that there is no required overheaad,
$\LEFTcircle$ to indicate minimal overhead, and $\CIRCLE$ to indicate that the
server would have additional requirements at least at
the same scale as it does during normal operation (i.e., its requirements would
at least be doubled).

\item[Trust assumptions:] We consider the assumptions needed for the
protocol to satisfy both privacy and integrity, in terms of which participants
have to trust which other participants to be sure that the correct information
is communicated in a privacy-preserving way.  We use `none' if there are no trust
assumptions, and otherwise describe them.

\item[Near-term deployability:] We consider how possible it would be to
deploy the protocol within the next 2-3 years.  This factors in both costs 
and trust assumptions, in terms of whether or not there are natural
participants in the \ctshort ecosystem who can play these roles.  We use
$\Circle$ to indicate that there are major obstacles, 
$\LEFTcircle$ to indicate significant but not insurmountable obstacles, and
$\CIRCLE$ to indicate that near-term deployability is a reasonable
expectation.

\end{sarahlist}

\subsection{Proposals for querying}

We first describe protocols for the querying phase, in which the goal is for a
client to find out from a log operator whether or not a specific entry is included
in the log without the log being able to link that specific entry to a specific
client.

\subsubsection{Network-level anonymization}\label{sec:anonymous-queries}

We first discuss proposals in which the client provides the queried
certificate/SCT in the clear, but their identity may be hidden from the log
at the network layer.

\paragraph{Query directly}\label{sec:direct-query}
The simplest proposal for querying a log is to have the client do so
directly; i.e., to request an inclusion proof from the log for a given
certificate.
This can be deployed easily, and similarly requires little overhead for an
individual client so is performant.  Following Section~\ref{sec:threat}, the fact
that the client receives an inclusion proof directly from the log means the
protocol achieves integrity.

The protocol does not achieve any privacy for the client, as it reveals
its certificates directly to the log.  There are two possibilities: first, a
client represents an individual browser, and the log thus learns
the website visited by that browser.  This achieves no privacy at all.  Second,
a client represents an auditor to whom a collection of individual
browsers have reported one or multiple certificates; e.g., a browser vendor with
whom some users have opted to share a portion of their browsing history.  In this
case, individual users achieve $k$-anonymity, where $k$ is the total number of
users of that auditor, but as they reveal their certificates directly to the
auditor they must trust it to not share them externally.  We discuss this
second case in more detail in Section~\ref{sec:phase1}.

\begin{table*}[!htbp]
\centering
\resizebox{\textwidth}{!}{%
\renewcommand\arraystretch{1.5}
\begin{tabular}{lccccp{1.73cm}cp{2.5cm}>{\centering\arraybackslash}p{1.5cm}}
& & \multicolumn{3}{c}{Client costs} & & & & \\
\cmidrule(lr){3-5}
Proposal & Privacy & Bandwidth & Storage & Computation & Certificate issuance
~~~~~~~~~~ latency & Server costs & Trust assumptions &
Near-term deployability \\
\midrule
Query directly (browser) & \Circle & \LEFTcircle & \Circle & \LEFTcircle & none
& \Circle & log & \CIRCLE \\
Query directly (auditor) & \LEFTcircle & \LEFTcircle & \Circle & \LEFTcircle &
    none & \Circle & auditor (depending on reporting phase) & \CIRCLE \\
Proxy/mixnet & \LEFTcircle & \LEFTcircle & \Circle & \LEFTcircle & none &
    \Circle & no collusion between proxy/mixnet and log & \CIRCLE \\
DNS & \LEFTcircle & \LEFTcircle & \Circle & \LEFTcircle & none & \Circle &
    no collusion between DNS resolvers and log & \CIRCLE \\
Fuzzy ranges & \LEFTcircle & \LEFTcircle~--~\CIRCLE & \Circle &
    \LEFTcircle~--~\CIRCLE & wait for sequencing & \LEFTcircle & none & \LEFTcircle \\
PIR & \hspace{4pt}\CIRCLE$^\caveat$ & \LEFTcircle & \Circle & \LEFTcircle &
    wait for sequencing & \CIRCLE & no collusion between replicated logs & \Circle \\
C3-PSM (log) & \hspace{4pt}\LEFTcircle$^\caveat$ & \LEFTcircle~--~\CIRCLE &
    \Circle & \LEFTcircle~--~\CIRCLE & none & \CIRCLE & none & \LEFTcircle \\
C3-PSM (third party) & \LEFTcircle & \LEFTcircle~--~\CIRCLE & \Circle &
    \LEFTcircle~--~\CIRCLE & none & \Circle & third party (for integrity) & \LEFTcircle \\
Local mirroring & \CIRCLE & \CIRCLE & \CIRCLE & \CIRCLE & none & \LEFTcircle &
    none & \CIRCLE \\
Fast embedding & \hspace{4pt}\LEFTcircle$^\caveat$ & \LEFTcircle & \Circle &
    \LEFTcircle & wait for inclusion & \Circle & none & \LEFTcircle \\
Slow embedding & \hspace{4pt}\LEFTcircle$^\caveat$ & \LEFTcircle & \Circle &
    \LEFTcircle & wait for MMD & \Circle & none & \LEFTcircle \\
OCSP stapling & \hspace{4pt}\LEFTcircle$^\caveat$ & \LEFTcircle & \Circle &
    \LEFTcircle & none & \Circle & none & \Circle \\
\cdashlinelr{1-9}
Report directly & \Circle & \LEFTcircle & \Circle & \LEFTcircle & -- &
    \LEFTcircle & auditor & \CIRCLE \\
Proxy/mixnet & \LEFTcircle & \LEFTcircle & \Circle & \LEFTcircle & -- &
    \LEFTcircle & no collusion between proxy/mixnet and auditor & \CIRCLE \\
Web server & \CIRCLE & \LEFTcircle & \Circle & \Circle & -- & \LEFTcircle &
    no persistent MitM attack; website will report & \Circle \\
C3-PSM & \hspace{4pt}\LEFTcircle$^*$ & \LEFTcircle~--~\CIRCLE & \Circle &
    \LEFTcircle~--~\CIRCLE & -- & \CIRCLE & none & \CIRCLE \\
ZKP of non-inclusion & \CIRCLE & \CIRCLE & \Circle & \CIRCLE & -- & \LEFTcircle &
    none & \LEFTcircle \\
\bottomrule
\end{tabular}%
}
\caption{Proposals for querying the log and, below the dashed line, for reporting
to an auditor.  Privacy is measured in terms of the difficulty of linking a specific
client to a specific entry.  The $\caveat$ superscript indicates that the
protocol achieves this level of privacy only with respect to a covert
adversary~\cite{aumann07tcc} rather than one that is fully malicious, and the
$*$ superscript indicates that it achieves this level of privacy only with
respect to an honest-but-curious adversary.  The \LEFTcircle~--~\CIRCLE~range
indicates that there is a tunable parameter that can increase or decrease the
overhead (but, as we discuss in the respective sections for these proposals,
this overhead is proportional to their privacy guarantees).}
\label{tab:query}
\end{table*}

\paragraph{Proxy/mixnet}\label{sec:proxy-query}

Rather than have each client contact the log directly, clients could
route their queries through a single proxy server or a series of proxies; i.e.,
a mixnet, as mentioned by Eskandarian et
al.~\cite{ct-privacy} and as used implicitly in the CTor protocol for Tor
clients due to Dahlberg et al.~\cite{dahlberg21pets}.

This protocol is performant and could be deployed in the near term; e.g.,
browsers could act as a proxy for \ctshort queries just as
some of them currently act as a proxy
for Safe Browsing queries~\cite{brave-proxy, apple-proxy}.  As in the previous
proposal, it achieves
integrity as the client receives an inclusion proof from the log.  It protects
the privacy of the client as
long as there is sufficient traffic, as clients achieve $k$-anonymity with
respect to the set of clients using the proxy at a given point in time.  This
assumes, however, that the proxy servers are not colluding with the log,
which is a problem for companies like Google that both offer a browser and run
CT logs.
Using more proxies makes it less likely that they are all colluding, but adds
latency to the querying protocol.

\paragraph{DNS}\label{sec:dns-query}
Using a proxy improves privacy by avoiding direct communication between the
client and the log.  Instead of adding an external proxy server, we could identify
a party who already knows of the client's interest in a given
certificate, and then route the query through them.  One such party is a DNS
resolver.
In 2015, Google proposed a DNS-based protocol for fetching inclusion
proofs~\cite{ct-chrome-design-doc,ct-over-dns-rfc}.  Briefly, the protocol
involves the client requesting records for domain names that encode
information about leaf hashes.  These special DNS records
then provide inclusion proofs for these leaf hashes.

This protocol is performant and could be deployed in the near term; furthermore,
the fact that the client receives back inclusion proofs means it satisfies
integrity.  In theory, the protocol also preserves the privacy of clients, who
already reveal the domain names they are interested in to their local DNS
resolvers; furthermore, the logs see the request in the clear but it comes from
their configured DNS resolver, which should reveal nothing about the client.
In practice, however, there are known
privacy risks associated with this approach~\cite{sct-dns-privacy}.  For
example, SCT auditing is done asynchronously, so a client might send the SCT
query hours after they actually visit the site.  This might cause their query
to be routed through a different DNS resolver, and thus reveal information.
Even with the same resolver, DNS resolvers may do a form of prefetching; i.e.,
resolving domain names linked to on the site that the client is
currently visiting.  They thus know only the domain names that they have
resolved for the client, so requesting an inclusion proof provides the
additional information that the client is not only resolving but actually
visiting a site.  As such, Google no longer seems to be focusing on this
protocol as a solution for SCT
auditing~\cite{no-more-dns}.

\subsubsection{Privacy-preserving queries}\label{sec:private-queries}

Rather than focus on anonymity, the next three proposals allow the
client's identity to be known to the log but use cryptographic techniques
to hide the specific certificate in which they are
interested.

\paragraph{Fuzzy ranges}\label{sec:fuzzy-query}
Instead of having a client query for a single leaf hash or index, they
could ask to see inclusion proofs for all entries in a range that they
know contains their specific certificate of interest.
This type of request is implicit in the work of Eskandarian et
al.~\cite{ct-privacy}.  To obtain this
range, one could imagine having a client query for either all certificates
within the range of a given timestamp, or for all entries between two indices in
the underlying data structure.  The former approach can be problematic for
privacy, as a client does not know how many certificates were logged in a given
time period, so may end up with a smaller anonymity set than desired.  More
importantly, this feature is not supported by the current CT
API~\cite{6962}.

The only currently available option for this type of query is thus the latter
option: having clients query for all entries between two indices.  This has the
upside for privacy that it fixes the size of the client's anonymity set,
but the downside that in order to identify the right range the
certificate would need to contain its index, or \emph{sequence number}.  This
adds latency into the certificate issuance process, which is a limitation we
discuss further in Section~\ref{sec:limit-latency}.  More generally, there is an
inverse relationship between performance and privacy: a client's communication and
computation costs are $O(k \log(N))$, so increasing the size of the
anonymity set means increasing these costs and thus degrading performance.
Furthermore, to avoid revealing the exact index $i$ of the certificate, a client
would need to add some random offset; i.e., they would
query for the range $[i+r-\frac{k}{2}, i+r+\frac{k}{2})$ for some random value
$-\frac{k}{2} <  r \leq \frac{k}{2}$.  In the (likely) case where the client queried
multiple logs on the same certificate, however, if those logs were colluding then
they would be able to perform an intersection attack~\cite{serjantov02pets} to
identify the certificate shared by both ranges, or at least to significantly reduce
the set of candidate certificates.

\paragraph{Private information retrieval (PIR)}\label{sec:pir-query}

Private information retrieval (PIR) aims to achieve a notion of
\emph{client privacy} in
which an adversarial server learns no information about the queries made by
clients.  PIR solutions do not typically achieve any notion of privacy for the
server, but
this is not needed for \ctshort as the contents of all logs are designed to be
globally visible.  In \ctshort, the value returned to the client is an
inclusion proof for their specific entry of interest.

Lueks and Goldberg were the first to propose using PIR for
CT~\cite{ct-pir}, and a more performant solution was later proposed by Kales,
Omolola, and Ramacher~\cite{ct-pir2}.  Recently, Kogan and
Corrigan-Gibbs proposed a PIR solution, Checklist, for the related problem of Safe
Browsing~\cite{kogan21usenix}.  The first two solutions are ``traditional'' PIR
protocols, while Checklist is an example of offline/online PIR~\cite{corrigan20eurocrypt}.
To avoid the high performance overhead of
single-server PIR, all three solutions operate in the two-server PIR model; i.e.,
they require two non-colluding servers to run identical copies of the log.

All three protocols provably achieve privacy for the client in retrieving a
record from the database.  For the \ctshort-specific solutions, this record
consists of a certificate and an inclusion proof, which means the protocol
satisfies integrity as long as the PIR database is the only one maintained by
the log.  Providing an inclusion proof, however, opens the protocol up to the
following attack by a fully malicious log~\cite{ayer18auditing}.  Because an
inclusion proof is formed
with respect to a specific STH, in addition to a specific entry
in the database, a malicious log could use a unique STH for each inclusion proof;
this would create a one-to-one mapping between STHs and certificates.  A client
wanting to verify inclusion of a certificate would need to not only verify the
inclusion proof returned by the log but also verify a consistency proof between
the STH used for the inclusion proof and one that they already know and trust.
If they query the log directly for this consistency proof, they reveal
the STH and thus reveal the certificate it represents.  We defer further
discussion of this issue until Section~\ref{sec:limit-privacy}, but briefly
mention here that without an additional
privacy-preserving method for retrieving consistency proofs this means
privacy can be achieved only with respect to a \emph{covert}
adversary rather than one that is fully malicious~\cite{aumann07tcc}.

In terms of performance, Lueks and
Goldberg evaluate their protocol on a 3\si{\giga\byte} database, which they argue
can store inclusion proofs for a log of 4~million
certificates.  Kales et al.\ evaluate their protocol for logs containing $2^{28}$
(268M) certificates, which as we see in Section~\ref{sec:discussion} is in line
with the size of many \ctshort logs today.  They find
that the server's computation on a query takes roughly
1~second, the work for the client takes under a millisecond, and the protocol
requires 6\si{\kilo\byte} in communication costs.  They also evaluate the work of
Lueks and Goldberg and find that it requires 3.5~seconds for the server to respond
to a query and 625\si{\kilo\byte} in communication costs.  Kogan and
Corrigan-Gibbs evaluate Checklist on a database of size 3~million (in line
with the smaller number of records used in Safe Browsing) and find that the
offline phase takes 11\si{\second} on both the
client and the server, requiring roughly 10\si{\mega\byte} in communication
costs.  In contrast, the online phase takes at most
1\si{\milli\second} for both the server and the client and requires roughly
1\si{\kilo\byte} in communication costs.
None of these costs are prohibitive given that SCT auditing is designed to be
performed asynchronously, although of course further investigation would be needed
to fully assess their practicality.

In terms of deployability, both \ctshort-specific solutions require a
certificate to contain its index in the log.  As discussed in
Section~\ref{sec:fuzzy-query}, this adds latency to the certificate issuance
process, which is a limitation we discuss in Section~\ref{sec:limit-latency}.
For all three solutions, there are significant deployment challenges
associated with keeping two servers fully synchronized.  Kales et al.\ suggest
that the second log could be hosted on a cloud platform run by a competitor of the
first log operator, but this would incur a significant financial cost.

\paragraph{Private set membership (PSM)}\label{sec:psm-query}

Private set membership (PSM) allows a client to learn whether or not an entry is
in a list held by a server, with the client learning nothing about the list
beyond (non)membership of the queried entry and the server learning nothing
about the client's entry. (This
differentiates PSM from PIR, which does not try to provide server privacy.) This
is a specific case of private set intersection (PSI) where the size
of the client's set is one.

We are not aware of any research using PSM for SCT
auditing, but previous research has explored its usage in the related context of
checking for compromised credentials (C3)~\cite{li19ccs,thomas19usenix,hibp,
apple}.
In these ``C3-PSM'' protocols, a user sends a prefix of the hash of their username
(which is less privacy-sensitive than their password) and the server store lists
of either credentials or password hashes in \emph{buckets} according to their hash prefix.  When a user
queries on a prefix, the server can thus send back the relevant bucket and the
user can check locally if their particular entry is in the bucket or not.

If we adapt this approach to CT, then each log would need to sort its certificates
into buckets, according to a prefix of the certificate's
hash.
If the
client simply ran the PSM protocol with the log, however, the overall protocol
would not satisfy integrity, as a malicious log might store certificates in the
buckets that are not actually in the log.  To prove that it wasn't doing this
and satisfy integrity, the log would also need to store associated inclusion
proofs (with respect to a given STH) for each certificate in a bucket as well.
This augmented protocol would thus impose a significant overhead on the
log in the form of both storing all the buckets (which, with the associated
inclusion proofs, use $N \log(N)$ space where $N$ is the number of
certificates in the log), and periodically recomputing an inclusion proof for
every certificate and updating the buckets accordingly.

In C3 protocols, however, the client is assumed to be interacting with a
party (such as their browser vendor) that they trust to maintain a complete list
of password hashes.
A more realistic solution here would thus be to adopt the same threat model as
for C3: a client would trust their browser vendor or some other third party to
maintain a complete list of certificates, and would engage in the PSM protocol
with them instead of the log.  The fact that the client trusts this third party
for integrity removes the need for inclusion proofs and thus means the client
can receive a simple yes/no response (as in a traditional PSM protocol).

The protocol between the client and the log achieves
$k$-anonymity, where $k$ is the size of the bucket, only if
the log is covert rather than malicious; this
is due to the same potential for a fully malicious log to use unique STHs for
inclusion proofs that we discussed for PIR.  Similarly, the protocol between the
client and a semi-trusted third party also achieves $k$-anonymity.
In either case, performance
is inversely proportional to privacy: revealing larger prefixes results in smaller
buckets, which means lower bandwidth overhead but a smaller anonymity set.
While the protocol achieves $k$-anonymity for a single query, it is
unclear how privacy would degrade over
repeated querying on potentially related certificates, such as certificates for
domains that fit into a common category~\cite{olejnik12pets,bird20soups}.
This may be an interesting and fruitful area for future research.

\subsubsection{Avoiding client querying}\label{sec:no-queries}

Any solution that requires communication between the client and log imposes some
performance overhead.
In theory, avoiding having any querying protocol at all would have
a positive impact on bandwidth (as the client and log do not need to
communicate) and on privacy, as the client cannot reveal any information
about the certificate if it does not perform any queries about it.  In
practice, these ``non-interactive'' protocols are limited in the privacy they can
achieve due to the
same limitation discussed in Sections~\ref{sec:pir-query}
and~\ref{sec:psm-query}: inclusion proofs are tied not only to specific entry
but also to a specific STH, which a malicious log may exploit.
We discuss this limitation further in Section~\ref{sec:limit-privacy}.

\paragraph{Local mirroring}\label{sec:mirror-query}

If clients mirrored every log, then they could check
for inclusion locally and would never need to issue any queries.  To some
extent, this solution resembles maintaining a certificate revocation list (CRL).

While this solution achieves integrity and perfect privacy, it clearly imposes a
significant storage requirement on the client, as there are billions of
certificates stored across all CT logs; even CRLs, which are much smaller, are
considered impractical for this reason.  The bandwidth overhead would be even
higher, as they would need to periodically update the list of entries they
maintain for each log.  Nevertheless, mirrors do exist today (for individual logs)
so such a solution would be relatively easy to deploy in the short term.

\paragraph{Embedding, fast and slow}\label{sec:embed-query}
RFC~6962-bis briefly suggests that inclusion proofs can be embedded in a
certificate via a custom extension~\cite[Section 7.1.2]{6962bis}.  This
leaves open several questions, however, one of which is who should be
responsible for embedding the inclusion proof.  The natural answer would
seem to be certificate authorities (CAs), given the inclusion proofs by the
logs, as otherwise this would require a mandatory change in web servers.

The next question is when CAs should obtain these inclusion proofs.  If they do
so as soon as the certificate gets included in the log (\emph{fast} embedding),
this adds latency to the process
of issuing a certificate, as CAs have to wait until it is included in
the log.  The effect on privacy is also negative, due to the fact that
the STH used in the inclusion proof is likely to represent a small number of
certificates even in the case that the log is honest; in the case that the log
is malicious, it can guarantee that the STH is unique to this certificate.
Thus, even in the honest case this means that the protocol achieves $k$-anonymity
for a relatively small value of $k$.

If instead CAs wait for some period to obtain the inclusion proof (\emph{slow}
embedding), such as the maximum merge delay (MMD),
the latency involved in issuing a certificate increases significantly.  In
terms of privacy, this achieves $k$-anonymity for a larger $k$.
We discuss in Section~\ref{sec:limit-privacy} the respective values of $k$ that
could be expected for these solutions, and discuss in
Section~\ref{sec:limit-latency} the challenges that they present for
deployability.

\paragraph{OCSP stapling}\label{sec:ocsp-query}
RFC~6962-bis also briefly mentions serving inclusion proofs using OCSP
stapling~\cite[Section 7.1.1]{6962bis}, as described in
Section~\ref{sec:revocation}.
Here, the web server periodically requests not only the certificate status from
a CA but also inclusion proofs for its certificates,
and serves this signed information to the client alongside the certificate
and its embedded SCTs.
To achieve integrity, this additional information needs to be required and
signed by the CA for all certificates, or adversarial web servers can simply
choose to not provide anything.

This approach seemingly avoids the privacy issues raised by
embedding an inclusion proof directly into the certificate: the only way to link
a client to their requested certificate is for the CA to maintain a mapping
from certificates to the STHs used for their inclusion proofs, and a colluding
log to maintain a mapping from IP addresses to the STHs queried for
consistency proofs; the combination of these two maps links a client
directly to a certificate.  While this level of collusion may
seem far-fetched, as we saw in Table~\ref{tab:basics} most logs are already run
by CAs.  If CAs and logs are colluding, this solution achieves
$k$-anonymity, where $k$ is the number of certificates that the CA gives out for a
given STH.  As discussed previously, this means that if we model CAs and logs as
fully malicious we cannot guarantee any privacy, so must instead treat
them as covert adversaries only.

The protocol is efficient for all parties.  In terms of
deployability, the main change required is at the CA rather than individual web
servers, but it does assume that web servers are set up to use OCSP
stapling.  We discuss the limitation of relying on changes in web servers
further in Section~\ref{sec:limit-changes}; briefly, Liu et al.\ found in
2015 that at most 5\% of certificates
were served by hosts that supported OCSP stapling~\cite{liu15imc}, and Scheitle
et al.\ found in 2018 that 0.01\% of their observed connections had SCTs served
via OCSP stapling.  Furthermore, in order for this proposal to achieve
integrity, web servers \emph{must} provide this information.  This means
certificates must suppport OCSP ``Must-Staple''; i.e., a version of OCSP
stapling in which responses are not considered valid if they do not contain the
requested information.  Chung et al.\ found in 2018 that only 0.02\% of
certificates supported this~\cite{chung18imc}.

\subsection{Proposals for reporting}

We now describe protocols for the reporting phase, in which the goal is for
an auditor to learn about entries that were not included in the log
without being able to link a specific reported entry to a specific
client.

\subsubsection{Network-level anonymization}

As in the querying phase, these proposals make no effort to hide the
client's certificate, but instead may hide their identity from the auditor at the
network layer.

\paragraph{Report directly}\label{sec:direct-report}

As in the querying phase, the simplest approach for reporting is to just have
clients send certificates to the auditor directly.  These can be
certificates for which the client has already performed the querying phase and
found to not be in the log, or they can be all or some subset of their
certificates, in which case the auditor can then perform the querying phase
itself (in this latter case, the client must trust the auditor for integrity).
Nordberg et al. mention the possibility of having individual browsers
send SCTs directly to trusted auditors~\cite[Section 8.3]{gossiping-in-ct}.

This solution requires very little overhead for an individual client, but
clearly requires them to trust the auditor fully as it does not achieve
any privacy.  We discuss in Section~\ref{sec:phase1} how such trusted auditors
may already exist for certain clients.

\paragraph{Proxy/mixnet}\label{sec:proxy-report}

As in the querying phase, a natural attempt to improve privacy would be
to have clients route their traffic through a single proxy or a
series of proxies, rather than contact the auditor directly.  Again, clients
could either send only certificates for which they have already performed the
querying phase and found to not be in the log, or all or some subset of their
certificates.  As with the previous solution, clients who do the latter must
trust the auditor to perform the querying itself.  This latter approach is used
by Dahlberg et al.\ in their CTor protocol for Tor
clients~\cite{dahlberg21pets}.

This solution has largely the same properties as the one in the querying phase:
it is performant and could be deployed in the near term by browsers that
already act as a proxy for the related problem of interacting with
Safe Browsing endpoints.  It also protects the privacy of the client as long as
there is sufficient traffic and as long as the proxy servers are not colluding
with the auditor.  Unlike in the querying phase, however, there may not be a high
volume of traffic going to the auditor.  In particular, the first scenario has
clients send only certificates that they have already determined are not
in the log.  We expect logs to violate the promise implicit in their
SCTs with very low frequency, given the serious consequences if they are caught
doing this, so in this scenario clients might need to
periodically send cover traffic, as they do in other mixnet
solutions~\cite{cai12ccs}.

\paragraph{Web servers}\label{sec:sct-feedback-report}

Nordberg et al.~\cite[Section 8.1]{gossiping-in-ct} proposed having a browser
report the SCTs that are relevant to a website it
is currently visiting.  It is then the responsibility of the web server
to send these SCTs to an auditor.

This method of reporting has the advantage that it is hard to disrupt without
also disrupting web browsing, and it preserves privacy as the web
server already knows the browser is visiting the site.  It assumes, however, that
an attacker cannot run a persistent man-in-the-middle attack, as they say
that ``clients will
    send the same SCTs and chains to a server multiple times with the assumption
    that any man-in-the-middle attack eventually will cease, and an honest server
    will eventually receive collected malicious SCTs and certificate chains.''
Even without such an attack, integrity also relies on the web server honestly
reporting their SCTs to an auditor, as a malicious web server could just decide
not to report and there would be no way to detect that they hadn't.  More
generally, it would require a change in a
significant fraction of web servers in order to capture certificate misissuance
at a broad scale.  We discuss this limitation further in
Section~\ref{sec:limit-changes}.

\subsubsection{Privacy-preserving reporting}

As with the analogous querying proposals, the next set of solutions allows the
client's identity to be known to the auditor but uses cryptographic techniques
to hide the specific certificate being reported (or,
for the first proposal, to hide it in all but exceptional cases).

\paragraph{Private set membership (PSM)}\label{sec:psm-report}

Google recently proposed a privacy-focused solution for having browsers report
certificates to auditors~\cite{opt-out}.  This protocol imagines that
an auditor (in their case Google) acts as a mirror for all \ctshort logs and thus
maintains a comprehensive list of all certificates; in the
maintenance of this list the auditor plays a role analogous to a Safe Browsing
endpoint maintaining a blocklist.
Before interacting with the auditor, clients first use a sampling strategy to
decide on a subset of certificates that they might report.  This step eliminates
interaction with the auditor in a manner similar to the Bloom filter used in
Safe Browsing; as mentioned in Section~\ref{sec:safe-browsing}, using a Bloom
filter is not practical here due to the significantly larger size of the list
and the fact that clients are looking for a missing rather than a matching
entry.
The client and the auditor then engage in a PSM protocol for each of the
certificates in this sample, analogous to the protocol used to check for
compromised credentials (C3).
If at the end of the protocol the client is convinced that its certificate is in
the list held by the auditor, they do not continue further.  If not, the client
reports the certificate to the auditor in the clear; i.e., sends it to
the auditor directly.  The protocol thus consists of a PSM ``querying'' phase
(but with the auditor rather than a log operator) followed by a direct reporting
phase.

In terms of privacy, the PSM phase of this interaction achieves $k$-anonymity,
where $k$ is the size of a bucket.  If the certificate is not
on the auditor's list the client reveals it directly to them, however,
which means the overall protocol achieves no privacy.  The proposal
suggests that this case is unlikely to happen in practice ``as Google maintains a
comprehensive copy of all valid SCTs'' and that in these rare cases it is thus
``appropriate to break anonymity.''   This assumes that the auditor honestly
maintains a comprehensive list of SCTs/certificates, which means the protocol
achieves $k$-anonymity only if the auditor is honest-but-curious.
Finally, as in Section~\ref{sec:psm-query}, the protocol achieves $k$-anonymity
for a single
query but it is unclear how privacy would degrade over repeated querying, and in
particular if the sampling strategy used in this proposal would resolve the
privacy issues present in Safe Browsing~\cite{gerbet15,kogan21usenix}.
Furthermore, performance is inversely proportional to privacy, as achieving
$k$-anonymity for larger values of $k$ requires sending larger buckets.

While an honest-but-curious adversary may make sense anyway when modelling the
role of a browser vendor, the proposal does contain additional mitigations.  For
example, it suggests that clients ``maintain a strict limit of 3 total SCT
reports'' sent to the auditor, which means that even a malicious auditor could
only ever see three certificates per client.

\paragraph{Proof of non-inclusion}\label{sec:zkp-report}

Rather than provide a non-included certificate directly to an auditor, a
client could instead provide a zero-knowledge proof of
its non-inclusion, as proposed by Eskandarian et al.~\cite{ct-privacy}.  In
other words, the client can prove knowledge of an SCT such that the timestamp
falls (strictly) between those of two adjacent log entries.

This protocol achieves provable zero knowledge, which means it is private, and
does not require any trust assumptions.  It does not fully satisfy integrity,
however, as the auditor learns only about the existence of a non-included entry.
If multiple clients provide such a report for the same log, the auditor does not
know if these reports represent the same certificate or repeated misbehavior and
is not in a position to follow up directly with the log and find out.
The auditor thus has no actionable evidence that it can use to further
investigate the potential misbehavior, which is a limitation we discuss further
in Section~\ref{sec:limit-reporting}.
In terms of performance, producing a proof requires over
five seconds on the client side (on a laptop) and the proof itself is over
333\si{\kilo\byte}.
As Eskandarian et al.\ argue, however, the reporting
protocol is likely to be run infrequently.

\section{Full Proposals}\label{sec:proposals}

In this section, we describe \emph{full} proposals for SCT auditing; i.e.,
proposals that combine a querying and a reporting protocol.  As compared with
the many individual components described in the previous section, there are
relatively few of these.  As we discuss further in
Section~\ref{sec:limit-reporting}, this is perhaps due to the fact that most
existing research has focused on the querying phase and not the
reporting phase.  We also see how a proposal in one phase can have
weaker privacy as a full protocol due to mismatched trust assumptions and
privacy guarantees in the other phase.

\subsection{Proofs of non-inclusion}

Eskandarian et al.\ proposed a protocol~\cite{ct-privacy} that combines fuzzy
ranges (Section~\ref{sec:fuzzy-query}) with proofs of non-inclusion
(Section~\ref{sec:zkp-report}).  Briefly, this means the client (a browser)
queries the log for a range that should include the certificate they are
interested in and then checks if their
certificate is indeed in this range.  If not, they provide a zero-knowledge
proof of its non-inclusion to some publicly accessible auditor.

The protocol as a whole (provably) preserves a client's privacy with respect to
the auditor but achieves only $k$-anonymity with respect to the log.
Furthermore, a client's queries to different logs likely reveal patterns that
shrink the anonymity set over time.
In terms of performance, there is a
significant bandwidth overhead for logs (who would be contacted by every
individual browser), and reporting a zero-knowledge proof imposes
significant costs in terms of both computation and bandwidth on the client but
is expected to happen infrequently (only in the case of non-inclusion).
Finally, in terms of functionality the protocol is not actionable in that the
auditor knows only that a log has misbehaved but not where or how many times.
We discuss this final limitation in more detail in
Section~\ref{sec:limit-reporting}.

\subsection{SCT Feedback}

SCT Feedback~\cite[Section 8.1]{gossiping-in-ct}
combines direct querying, between
an auditor and a log, with the reporting mechanism described in
Section~\ref{sec:sct-feedback-report}.  Briefly, this means the client (a
browser) reports the SCTs that are relevant to a website it is currently
visiting.  The web server then collects these SCTs and passes them on to some
publicly accessible auditor, who in turns queries the log for their inclusion.

The protocol as a whole preserves the privacy of the client, who reveals
their SCTs only to a website they are already visiting.  As discussed above,
however, the protocol has the significant downside that it relies on websites
to report certificates to an auditor.  This means that clients may not be protected
against malicious websites, and more generally that a
significant fraction of web servers would be required to run the protocol in
order to have a reasonable chance of identifying misbehavior.  We discuss this
limitation further in Section~\ref{sec:limit-changes}.

\subsection{Opt-in SCT auditing}\label{sec:phase1}

One active proposal by Google allows clients to \emph{opt in} to SCT
auditing~\cite{opt-in}; in fact, this has been deployed in Chrome as
of March 2021~\cite{ct-2021-plans}.  This combines direct querying
(Section~\ref{sec:direct-query}), between the
auditor and the log, with direct reporting (Section~\ref{sec:direct-report}),
from the client to the auditor.  The auditor in this case is Google.

While direct reporting raises obvious privacy concerns, the clients who report
SCTs already share their browsing history with Google
by performing extended reporting as part of Safe Browsing.  Furthermore, the
proposal states that ``Third-party logs don’t receive any information about
Chrome users' browsing history because we query Google-operated mirrors of CT
data instead of querying the logs directly.''  Thus, as the auditor sits within
the same trust boundary as the log, no information about the client's data is
revealed to anyone except the auditor.

\section{Discussion}\label{sec:discussion}

In this section, we discuss the limitations of existing solutions, in terms of
the assumptions they make and the requirements they impose.  The particular issues
we highlight are:
(1) certificate issuance latency,
(2) client constraints,
(3) privacy,
(4) significant changes to the Web infrastructure, and
(5) reporting misbehavior to an auditor.
We use these limitations to define a set of constraints for, and thus a clear
definition of, the problem of performing SCT auditing in the existing \ct
ecosystem.

\subsection{Issuance latency}\label{sec:limit-latency}
Logs may take up to 24 hours to include a certificate in a log, given the
current MMD for \ct, and as reported by Gustafsson et al.~\cite{gustafsson17pam}
it can take logs up to 12 hours to publish a new STH (which signals the
inclusion of a new batch of certificates).

Web hosting providers currently promise significantly faster issuance rates,
ranging from minutes to several hours~\cite{google-latency, godaddy-latency,
amazon-certs}.
Furthermore, issuing a certificate quickly is important to these
businesses, with ``slow certificate deployment
[leaving] customers with an unsatisfactory
experience''~\cite{cloudflare-for-saas}.  Shortening the MMD would address this
tension,
but would place a significantly higher
burden on log operators and---as we discuss below---could be harmful to privacy.
This higher burden would raise the barrier to entry for running a log, which
would ensure that only large institutions would be able to act as log operators
(as is already largely the case today).

It is thus infeasible for certificate authorities to wait for log inclusion
before issuing a certificate, which means \textbf{inclusion proofs cannot be
embedded into certificates}, as is required in the fast and slow embedding
proposals (Section~\ref{sec:embed-query}).  Furthermore, the current most widely
deployed log implementation, Trillian~\cite{trillian}, has no distinction
between sequencing an entry and including it.  This also means that at least
today, \textbf{sequence numbers cannot be
embedded into certificates}, as is required in the fuzzy ranges
(Section~\ref{sec:fuzzy-query}) and PIR (Section~\ref{sec:pir-query}) querying
proposals.

\subsection{Client constraints}
Clients are run on commodity laptops and, in an increasing majority of
cases,\footnote{\url{https://gs.statcounter.com/platform-market-share/desktop-mobile-tablet}}
on mobile devices.  These are computing environments that are limited in terms
of the bandwidth, storage, and computational power available to them.
Collectively, today's active time-sharded CT logs contain
5.8 billion certificates.  Even if a client stored only this many hashes (which
ignores storing all internal log hashes or the actual underlying data), this would
amount to 185.6\si{\giga\byte} of data, which means that
\textbf{clients cannot act as mirrors}, as is required in the local mirroring
querying proposal (Section~\ref{sec:mirror-query}).

\subsection{Privacy}\label{sec:limit-privacy}

Many querying proposals focus on privacy-preserving ways to retrieve inclusion
proofs, which are inherently tied to the certificates for which they prove
inclusion.  As
discussed first in Section~\ref{sec:pir-query}, however, an inclusion proof is
also tied to the STH with respect to which it proves inclusion, and for a client
to fully convince themselves that an entry really is in the log they need to
both (1) verify the inclusion proof for that entry with
respect to its associated STH and (2) verify a consistency proof between that
STH and one it currently holds and believes to be valid.
If an STH were used to form only one or a small number of inclusion proofs,
querying a log for a consistency proof with respect to this STH would reveal
significant information about the certificate.
As discussed earlier, it is possible for a malicious log to ensure that each
inclusion proof is formed with respect to a unique STH, thus enabling this
attack.

In the proposals that rely on network-level anonymization, this attack
is not effective as clients can use the same anonymous communication tools to
retrieve consistency proofs as they do to retrieve inclusion proofs.
It is not clear, however, how to modify the PIR (Section~\ref{sec:pir-query}),
PSM with the log (Section~\ref{sec:psm-query}), embedding
(Section~\ref{sec:embed-query}), and OCSP stapling
(Section~\ref{sec:ocsp-query}) querying proposals.
For these proposals, we must thus assume weaker \emph{covert
adversaries}~\cite{aumann07tcc}, who may deviate from the protocol but ``do not
wish to be `caught' doing
so.''  Given that monitors and auditors can both catch this form of log
misbehavior, we believe this is a reasonable way to model
adversaries in \ctshort.

Even honest-but-curious logs may still
unintentionally use a single STH to form only a small set of inclusion proofs.
To understand the extent to which this might be a problem today, we sought to
identify how many certificates are represented by a given STH.  In
particular, we performed two experiments as follows.

\begin{enumerate}

\item We queried each time-sharded log every
30 seconds for a week, starting on November 24, 2021, and included results for
all logs for which we observed at least 95\% availability.  This meant excluding
only the TrustAsia logs.

\begin{figure}[t]
\centering
\includegraphics[width=1.0\linewidth]{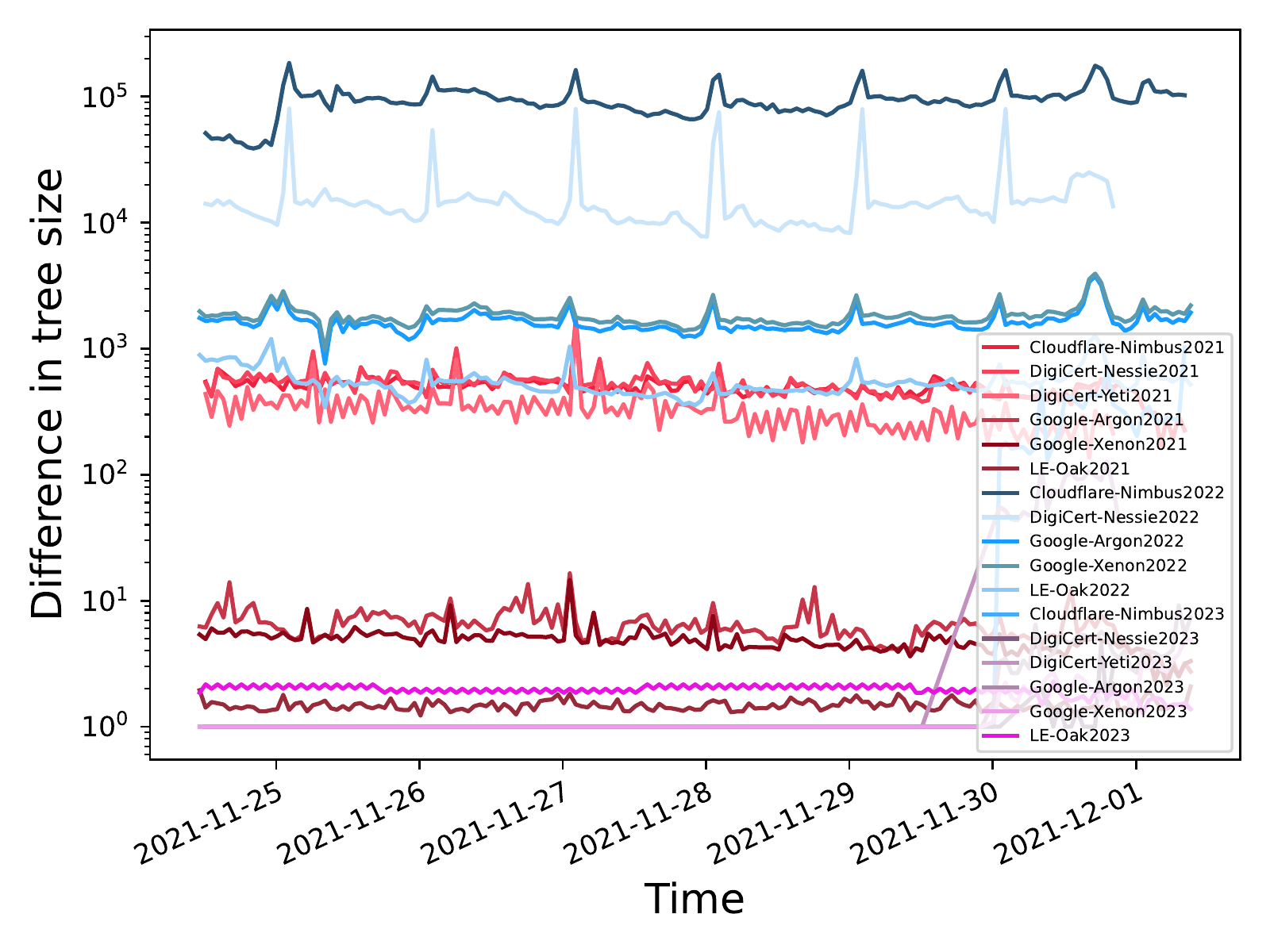}
\caption{The difference in tree sizes for time-sharded CT logs, averaged hourly
over a week (November 24--December 1, 2021) and at log scale.}
\label{fig:diff_sizes}
\end{figure}

\begin{table}
\centering
\begin{tabular}{lS[table-format=2.1]S[table-format=2.1]S[table-format=2.0]}
& \multicolumn{3}{c}{Log shard} \\
Log name & {2021} & {2022} & {2023} \\
\midrule
Cloudflare Nimbus & 1 & 1 & 1 \\
DigiCert Nessie & 1 & 1 & 1 \\
DigiCert Yeti & 1 & 1 & 1 \\
Google Argon & 22.8 & 60 & 3 \\
Google Xenon & 25.5 & 60 & 3 \\
Let's Encrypt Oak & 4 & 43 & 2 \\
\bottomrule
\end{tabular}
\caption{The number of unique STHs seen per minute for each log shard, averaged
across four runs of one query per second.  The runs were conducted between
November 22nd and 25th, 2021.
}
\label{tab:num-sths}
\end{table}

\item We queried each time-sharded log, this time once every second for a minute,
for a total of four minutes spread between November 22nd and 25th, 2021.

\end{enumerate}

The first experiment is designed to evaluate the case in which a log server
issues an inclusion proof with respect to the latest STH at the time a certificate
is included, as in the fast embedding approach (Section~\ref{sec:embed-query}).
This achieves $k$-anonymity, where $k$ is the difference in tree size between
that STH and the previous one.  As suggested by the results in
Figure~\ref{fig:diff_sizes}, there are many logs for which each STH represents a
fairly small number of entries: the 2022 log shards saw differences in
tree sizes in the hundreds or thousands, but for the 2023 log shards the average
difference was 23.6 and the median was 1.75. (Across all logs and shards, the
average difference was 8093 but the median was 385.)
For the 2023 log shards, even forming inclusion proofs
once per day would thus provide an anonymity
set of only 100-200 certificates.  Thus, \textbf{inclusion proofs cannot
be embedded into certificates}, as is suggested in both the fast and slow
embedding querying proposals (Section~\ref{sec:embed-query}).

The second experiment is designed to evaluate the case in which a CA queries a
log server for an inclusion proof at a given point in time, and again it
returns an inclusion proof formed with respect to the latest STH.  As suggested
by Table~\ref{tab:num-sths}, some logs form new STHs significantly faster than
once per minute, and potentially even faster than once per second.  As expected
given that we performed this experiment in November 2021 (in which the vast
majority of issued certificates would be expiring in 2022), this was especially
true for the 2022 log shards.  If we take
the Let's Encrypt Oak 2022 log shard as an example, in which
Figure~\ref{fig:diff_sizes} shows a difference in tree size of hundreds
between STHs sampled at a 30-second interval, this suggests that the anonymity
set for an STH retrieved at any given second would be at most tens of certificates.

Finally, if individual browser instances queried the log directly to retrieve
inclusion proofs then even honest-but-curious logs would be
able to learn the certificates seen by those clients.  Thus,
\textbf{browsers cannot directly query logs}, as is suggested by the first
option in the direct querying proposal (Section~\ref{sec:direct-query}).

\subsection{Significant Web changes}\label{sec:limit-changes}
Previous research has already shown that web servers are typically slow to adopt
new protocols, such as HTTPS~\cite{USENIX:FBKPBT17,mirian18tr,singanamalla20imc},
OCSP stapling~\cite{chung18imc}, newer versions of
TLS~\cite{kotzias18imc,holz20ccr}, and HSTS and HPKP~\cite{kranch15ndss}.
Furthermore, web servers are slow to patch even significant and
highly publicized security vulnerabilities, with a long tail never performing
any patching at all~\cite{durumeric14imc,li16usenix,stock16usenix}.

More specifically to CT, Gasser et al. explored adoption of the gossiping
endpoints proposed by SCT Feedback and found that at most $0.015\%$ of domains
had made them available~\cite{gasser18pam}.  Even if this number were higher,
even in a longer timescale it is not feasible to imagine every single web
server adopting a new protocol, so
\textbf{security cannot rely on mandatory changes implemented in web servers},
as is required in the OCSP stapling querying proposal
(Section~\ref{sec:ocsp-query}) and the SCT Feedback reporting proposal
(Section~\ref{sec:sct-feedback-report}).

\subsection{Reporting}\label{sec:limit-reporting}

The decisions that have thus far been made to remove logs from the \ctshort
ecosystem~\cite{aviator-policy2,startcom-policy2,wosign-policy2} have required
significant discussion~\cite{aviator-policy, startcom-policy, wosign-policy},
including a ``post-mortem'' from the log operators identifying the conditions
that led to their (unintended) misbehavior and considering how those conditions
could be prevented in the future.  As such, \textbf{auditors require actionable
and concrete evidence of a log's misbehavior}, as opposed to just learning about
its existence as in the ZKP of non-inclusion proposal
(Section~\ref{sec:zkp-report}).

More generally, as we saw in Section~\ref{sec:components}, much of the research
addressing the problem of SCT auditing has focused on the querying phase, with
significantly fewer proposals for the reporting phase.  While existing querying
proposals could perhaps be extended using existing reporting proposals, we
already saw in Section~\ref{sec:proposals} that the security of the full
proposal is only as secure as the weaker of the two phases.  Furthermore, if
we look at
the viable proposals remaining in Table~\ref{tab:query} we can see there are
few natural pairings.
The proposal for reporting directly requires full trust in the
auditor,
and the C3-PSM
proposal achieves privacy only with respect to an honest-but-curious auditor
(and in that case achieves $k$-anonymity rather than full privacy).

This leaves proxying as the only option, which pairs naturally with the
same proposal in the querying phase.  Indeed, the fact that Brave and Apple
already proxy queries to Safe Browsing endpoints suggests that this could be a
viable solution in the near term for these browsers, in terms of acting as a
proxy for both querying logs and reporting to auditors.  Chrome users, however,
comprise 65\% of all browser
users.\footnote{\url{https://gs.statcounter.com/browser-market-share}}
Given that Google acts as a CT log operator and is proposing to act as an
auditor, it would thus need another entity to act as a proxy, which is a
significant undertaking.
Moreover, if other browser vendors chose to act as auditors too
(which would of course be preferable to having Google be the only auditor),
they would also require third-party proxying solutions.

\subsection{Summary of constraints}

To summarize the constraints we have identified above, the main limitation of
many of the solutions we presented is that they address only the querying phase
and have no solution for reporting.  In particular, for almost all discussed
solutions for the querying phase, it is an individual browser instance that learns
whether or not a certificate is in the log, but there are both privacy and
feasibility questions in terms of how individual browsers could then report log
misbehavior to the broader CT ecosystem in a way that is actionable.
Furthermore, as discussed in Section~\ref{sec:limit-reporting}, most existing
solutions for querying are not compatible in terms of their threat model with the
existing solutions for reporting.  Thus, the main guideline for future proposals
is to consider both the querying and reporting phases and ensure that privacy is
preserved across both types of interactions.

In terms of privacy, our main observation in Section~\ref{sec:limit-privacy} is
that even for honest logs it is often possible for an STH itself to contain
significant information in terms of the number of log entries it represents.  It
is thus important for querying proposals to consider not only how clients can
preserve their privacy in obtaining inclusion proofs, but also how they can
do so in obtaining consistency proofs.  An alternative is to impose a rate
limit on the number of STHs that can be produced within a given period of time,
as is suggested in RFC~6962-bis~\cite[Section 4.10]{6962bis}.  This requires
substantial changes to log operation, however, which hinders near-term
deployability.  Furthermore, while limiting the number of available STHs would
improve privacy with respect to honest-but-curious logs, malicious logs could
still use specific STHs in the inclusion proofs for specific certificates to
perform fingerprinting.

\section{Conclusions}

In this paper, we have systematically explored the techniques that have
been proposed thus far for the problem of performing SCT auditing\dash which is
central to the security guarantees that \ct brings to the HTTPS
ecosystem\dash in a way that preserves the privacy of individual users.  In doing
so, we have identified proposals from both academia and industry, many of
which exist in quite different forms; e.g., posts on mailing lists, academic
papers, and readmes in GitHub repositories.  Despite these differences, we have
brought these proposals together and
explored them under a unified evaluation framework that considers their privacy,
integrity, performance overheads, trust assumptions, and near-term
deployability.

In doing so, we have identified several key limitations shared by multiple
proposals, in terms of (1) the increased latency they cause for certificate
issuance; (2) the excessive performance overheads they impose on clients; (3)
the limited privacy they achieve due to their ability to privately retrieve
inclusion proofs but not consistency proofs; (4) their need for change in a
significant majority of web servers in order to achieve integrity; and
(5) their lack of a proposed reporting component.   In highlighting these
limitations, our goal is to create a set of constraints that we hope serves as a
useful guide to researchers interested in this problem.  To further this goal, we
have also highlighted both the similarities and differences with other
problems associated with sensitive browsing-related information, most of which
have been explored and evaluated more thoroughly than SCT auditing has to date.

\section*{Acknowledgements}

This research received no specific grant from any funding agency in the public,
commercial, or not-for-profit sectors.

\end{document}